\documentclass{aa}
\usepackage{epsfig}

\begin{document}


   \title{Vela, its X-ray nebula, and the polarization of pulsar radiation}


   \author{V. Radhakrishnan \& A. A. Deshpande}

   \offprints{desh@rri.res.in}

   \institute{Raman Research Institute,
              C. V. Raman Avenue, Bangalore - 560 080, India\\
          \email{rad@rri.res.in; desh@rri.res.in}}

   \date{Received 19 July 2001 / Accepted ?}

   \abstract{
   The recent identification of the perpendicular mode of
   radio polarization as the primary one in the Vela pulsar by Lai et al.
   (2001) is interpreted in terms of the maser mechanism proposed by Luo \&
   Melrose (1995). We suggest that such a mechanism may also be operative for
   the parallel mode which opens up the possibility of accounting for all
   types of polarization observed in pulsars.
   We propose an alternative interpretation of the arcs in the nebular
   X-radiation observed by Pavlov et al. (2000) \& Helfand et al. (2001)
   with the Chandra Observatory, and interpreted by the latter as an
   equatorial wind. We interpret the arcs as traces of the particle
   beams from the two magnetic poles at the shock front. We also
   propose that the alignment with the rotation axis of the jet-like
   feature bisecting the arcs is an effect of projection on the
   sky plane and that there is no physical jet along the axis of
   rotation.
   }
   \authorrunning{Radhakrishnan \& Deshpande}
   \titlerunning{Vela, X-ray nebula \& pulsar polarization}
   \maketitle
   \section{Introduction}

Unlike in the case of most other non-thermal radio sources,
the polarization of the radiation from pulsars played an early and
fundamental role in attempts to understand and model the operative
emission mechanism. The high percentage of linear polarization, well over
the maximum theoretical limit for synchrotron radiation, together with a
special type of systematic sweep of the P.A. observed in the Vela Pulsar
led to the ``magnetic pole model" (Radhakrishnan \& Cooke 1969). The sweep
of the P.A. across the pulse was interpreted in terms of the line of sight
tangentially encountering different field lines close to the magnetic pole
as the pulsar rotated; and the parameters of the so called 'S' curve of
the P.A. sweep have ever since been interpreted in terms of $\alpha$ and
$\beta$, the
angles made by the magnetic axis to the rotational axis and to the line of
sight (at minimum impact angle) respectively. An important point is that
while the geometry of the 'S' curve is intimately related, through
$\alpha$ and $\beta$, to the locus of the sight line,
the actual angle between the plane of
polarization and the operative magnetic field line can have any value, as
long as it remains fixed. In the case of synchrotron radiation, the most
widespread emission mechanism invoked for non-thermal sources before the
discovery of pulsars, the electric vector of the radiation would be
perpendicular to the projected magnetic field, as the acceleration of the
charged particles was due to their gyration around the field lines.  

In the case of pulsars, the systematics of the polarization sweep, and its
independence of observing frequency, indicated clearly that the radiation
emanated from close to the polar cap in a region that had no internal
Faraday rotation. The strength of the fields associated with these regions
was so high that any transverse momentum and energy would be radiated away
``instantly", and the charged particles would be in their lowest Landau
levels and constrained to move along the magnetic field lines, like beads
on a string. An appreciation of this constraint led to the suggestion
(Radhakrishnan 1969) that the radiation could be due to the acceleration
in the plane of the curved field lines, and has been known since then as
``curvature radiation". As the motion of the particles, whether electrons
or positrons, could be only along the field lines, the polarization of the
emitted radiation should have the electric vector parallel to the
projected field lines. A consequence of this was the identification of the
intrinsic plane of polarization at the centre of the pulse (or more
correctly the inflexion point of the S curve), with the projection of the
rotation axis of the pulsar on the sky. This has had important
implications for a variety of studies over the years relating to the space
velocities of pulsars.  

According to the above picture, the PA of the
polarization can have one and only one value at any pulse longitude since
the angle of the projected field line is fixed. But as early as
1975  (Manchester et al. 1975; Backer et al. 1976) 
it was discovered that the PA could have more than one value
at a given longitude! Closer investigation revealed that the PA switched
between two modes, taking any one of two values which were orthogonal to
each other (Backer \& Rankin 1980). The polarization sweep pattern in any
one mode appeared identical to that in the other, barring the 90$^o$ shift in
P.A. There has been no shortage of attempted models for the radiation
mechanism, but in the absence of any other that could
be meaningfully compared with observations, the simple picture of the
magnetic pole model, with its rules for deriving $\alpha$ and $\beta$,
has survived for
over three decades, despite the blatant sweeping under the rug of the
observed freedom of the polarization vector to take one of two orthogonal
values, neither of which was ever shown to have a definite orientation
with respect to the field direction!!

\section{THE X-RAY VELA STORY}

 We turn now to a discussion of some observations
which appear to offer for the first time the possibility of establishing a
clear relationship between the directions of polarization and the magnetic
field of the pulsar. 

Recent observations of the Vela pulsar, and its
immediate surroundings, with the Chandra X-ray Observatory show a
two-sided jet at a position angle coinciding with that of the
proper motion of the pulsar (Pavlov et al. 2000; Helfand et al. 2001).
Lai et al.
(2001) have argued that the symmetric morphology of the X-ray emission
about the jet direction suggests strongly that the jet is along the spin
axis of the pulsar. As corroborating this interpretation, they cite
polarization observations using data from Deshpande et al. (1999) which
have been corrected for Faraday rotation both in the interstellar medium
and in the ionosphere. As the radio P.A. is at right angles to the
direction of the X-ray jet, Lai et al. (2001) conclude that the
polarization mode dominant in the Vela pulsar is one where the electric
field in radio emission is orthogonal to the magnetospheric field. 

Further
support for this picture comes from recent radio observations of the
region that have revealed a double lobe structure, with well separated
lobes of comparable intensity (Dodson et al. 2001; private communication).
These radio lobes are symmetrically placed on either side of the X-ray jet
with their diffuse inner edges very close to the boundaries of the X-ray
nebula, and independently suggest that the projection of the star's spin
axis must match the observed direction of the jet in the sky plane.

From
what has been discussed above, it must be concluded that the electric
vector of the Vela pulsar's radiation is perpendicular to the plane
containing the magnetic field-line,
and not parallel to it as one has generally assumed.
This was also noted by Helfand et al. (2001) although their major concern
was with the morphology of the X-ray nebula and its interpretation. We
shall also discuss the X-ray nebula shortly, but first the implications of
the polarization P.A.

\section{ IMPLICATIONS FOR THE EMISSION/AMPLIFICATION MECHANISM}

 There is no
question that the charged particles in the polar cap regions of any pulsar
will be constrained to move essentially along the field lines as already
mentioned earlier. There is also no question that there will be radiation
from these relativistic particles due to the acceleration associated with
the curvature of the field lines, and that the polarization of this
radiation will be linear and parallel to the projection of the field
lines, assumed planar for the moment. But the brightness temperature of
such radiation cannot exceed the kinetic temperature of the electrons (and
positrons), which for even extreme values of the magnetic field and spin
period are unlikely to be within orders of magnitude of the inferred
brightness temperatures, as has been well known since the earliest
observations. The absolute need for maser-like amplification, whatever the
mechanism of the input radiation, has thus always been recognised, and has
motivated numerous attempts over the decades to propose models for the
radiation mechanism of pulsars. 

As noted at the beginning, one of the
striking characteristics of pulsar radiation, which must be accounted for
in any theoretical model, is its polarization behaviour. Another, as just
seen, is the extremely high brightness temperature. We find it remarkable
that both these characteristics seem to be well accounted for in the model
put forward by Luo \& Melrose (1995). They assume that the input signal
is curvature radiation, which as discussed in detail above, seems
eminently reasonable to us. The amplification process requires a certain
non-planarity of the field lines, for which there has long been evidence
from many different lines of investigation on pulsars (Radhakrishnan 1992)
The surprising aspect is that while the spontaneous curvature radiation is
polarized parallel to the field lines, the amplified output is
perpendicular to it, in agreement with the observations discussed in the
last section. It would appear therefore that the ``normal" polarization
mode is really ``orthogonal" (to the field lines) for reasons associated
with the physics of the amplification process, without which pulsars would
not be detectable.

It should be pointed out that the above is not a
violation of the fundamental requirement that in any amplifier the
stimulated emission should be indistinguishable from the stimulating input
signal. Because of the torsion in the field geometry, the spontaneous
emission has a small component in the direction perpendicular to the field
in the amplifying region, and it is this component that, according to Luo
\& Melrose (1995), is preferentially amplified and dominates the output.

The pattern of switching to orthogonal modes of polarization varies from
pulsar to pulsar and can happen in different parts of the pulse profile
for different pulsars. For Vela, the existance of both modes was noticed
as early as 1983 by Krishnamohan \& Downs.
Although the radiation in the other mode is on
the average much weaker, it is important to appreciate that detectable
radiation in any mode, in any pulsar, always corresponds to brightness
temperatures that require enormous amplification. We are thus forced to
conclude that there must be more than one mode of amplification if the
input signal is the spontaneous ``curvature radiation", as is manifest by
the ``S" curve outputs of the amplifier, in whichever mode it is operating.
This implies that there must be conditions when the parallel mode develops
more negative absorption than the perpendicular one favoured by Luo \&
Melrose, and results in a polarization flip. The probability of this
happening could be influenced by the particular field distortions present
over the longitude range in question. But the fact that either mode can
occur is very reminiscent of maser processes in general, where different
allowable modes typically compete with each other resulting in one of them
rapidly taking over all of the available power.  

The observation in many
pulsars, of elliptically polarized radiation of detectable strength, is
further evidence of the existence of amplification in both modes, but now
simultaneously with some phase difference and different gains. A major
obstacle in the understanding of pulsar polarization until now has been
the difficulty of seeing how electric fields could be generated
perpendicular to the ultrastrong magnetic field lines, only along which
the charges were constrained to move. The mechanism of Luo \& Melrose
(1995) - thanks to torsion - allows and predicts radiation perpendicular
to the field lines, and now reduces the explanation of any type of
polarization in pulsars to a matter of detail, as opposed to a difficulty
of principle.

\section{THE X-RAY ARCS: TRACERS OF THE RADIATION BEAMS?}

 We turn now to the
spectacular X-ray image provided by Chandra with two remarkably
symmetrical arcs bisected by the jet like feature mentioned earlier.
Helfand et al. (2001) have put forward a detailed model where they ``assume
that the two arc-like features lie along circular rings highlighting
shocks in which the energy of an outflowing equatorial wind is dissipated
to become the source of synchrotron emission for the compact nebula" and
attribute the incompleteness of the rings to preferential Doppler boosting
of the emission in the forward direction. They also ``assume that the two
rings straddle the equator symmetrically and suppose that the deficit of
emission exactly in the equatorial plane is related to the fact that this
is where the direction of a toroidally wrapped magnetic field changes sign
i.e. the field may vanish there". They go on to derive the half opening
angle of the wind $\theta$ as 23$^o$.3, and the radius of the shock $r_s$
as a/dCos$\theta\sim 1\times10^{17}$ cm for d = 250 pc.

We would like to propose a somewhat
different model for the X-ray arcs, starting from the magnetic pole model
for pulsar radiation discussed at length earlier, and that is invariably
miscalled the ``rotating vector model\footnote{This term was introduced
by proponents of light cylinder emission models.
A rotating vector is what is observed; the essence of the R.C. model was
the location of the emission in the magnetic polar vicinity where the
observed rotation would occur naturally. Ironically, the term would be
perfectly appropriate for what we are proposing in this section and the
next.}".
In that model, the radiation (and
also its amplification as just seen) are produced by highly relativistic
particles streaming out along the open field lines from both magnetic
poles. We now examine the X-ray data for the possibility that the two arcs
reflect the traces of these two particle beams as they encounter
the ``walls" surrounding the central cavity created by the pulsar. Such a
cavity was elaborated in the classic paper by Rees \& Gunn (1974)
for the Crab, and has since formed a part of most, if not all, subsequent
discussions and models of pulsar created nebulae.

\begin{figure}
\epsfig{file=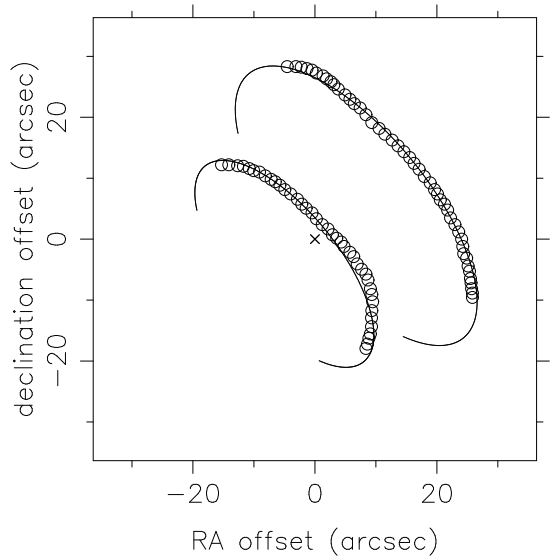,width=2.9in,angle=-90}
\caption[]{Our best fit model for the pair
of bright arcs seen in the Chandra Vela X-ray image.
The pulsar location (at 0,0) is denoted by a cross; 
the circles denote
the arcs seen in the X-ray image.
The continuous lines show our best fit model where these arcs 
are the locii of the rotating magnetic-axis vectors.
(see text for further details).}
\end{figure}

We assume that the particles leave the weakening field lines at some point
well before sweep-back effects set in close to the light cylinder, and
proceed ``ballistically" outwards. If this picture is valid, one of the two
arcs should pass close to our sight line to the pulsar, as indeed it does.
To assess this further, we have modelled the arcs as the near-side
portions of two rings (seen in projection) traced by sweeps of the
(magnetic) polar cones. The rotating magnetic-axis vector is as
described in Deshpande et al. (1999).
The model parameters are the pair of radial
distances ($r_1$ \& $r_2$ as measured from the star location and 
expressed in arcseconds)
associated with the two ring traces, 
the inclination ($\alpha$) of the magnetic
axis of the star to its rotation axis, the angle of closest approach
(the impact angle $\beta$) of the magnetic axis to our sight-line, and the
position angle (PAo) of the rotation axis projection on the sky-plane. The
angle ($\zeta$) between the rotation axis and our sight-line is simply
$(\alpha + \beta)$.
Desired consistency with radio polarization observations would allow only
certain combinations of the viewing geometry; that is $sin\alpha/sin\beta$
should be equal to the steepest sweep rate
($d\chi/d\phi$) of the polarization position
angle with respect to the rotational longitude. When we constrain the
$\alpha \& \beta$
combinations using $(d\chi/d\phi)_{max}$ 
of -9 degree/degree (as listed by Lyne \&
Manchester 1988), the best fit PA$_0$ is found to be 129 degrees (measured
from North through East), the radial distances $r_1$ \& $r_2$ 
are unequal (about 22 \& 29
arcsec for the near and the opposite polar cones respectively)
and $\alpha\sim$ 71 degrees ( $\beta\sim$-6 degrees correspondingly).
Note that this implies a value of $\zeta$ of about 65 degrees,
significantly larger than the 53
degrees estimated by Helfand et al. (2001) based on their model of an
equatorial torus. The above model is illustrated in Fig. 1.

The changing direction of the magnetic axis viewed in projection on the
sky plane as the star rotates is described in exactly the same way as that
for the position angle of the radio polarization (Deshpande et al. 1999)
whatever their relative difference. The proposed association of the arcs
with the traces of the polar emission beams thus provides a new and
independent means to probe the viewing geometry. They can sample (as they
do in the present case) a much larger fraction of the rotation cycle and
can provide additional constraints on the viewing geometry. One crucial
such constraint becomes available via the trace of the polar
emission from the other pole that is generally not available unless
interpulses are observable. Also, interestingly, the ``sign" of the impact
angle $\beta$ 
would become readily apparent from the beam traces even if they do
not sample large fractions of the rotation cycle, and without needing to
know the sense of rotation of the star.

A simple calculation shows that
the visible extent of the arcs is consistent with an X-radiation spread
confined to about 70 degrees around the respective directions to which the
magnetic axis points as the star rotates. We consider this as explained
simply in terms of the initial dispersion of the motions as the particle
beam ploughs into the region of compressed toroidal field - the shocked
region as usually described - and the consequent spread of the pitch angle
distribution. By just assuming that we will see radiation as long as there
is some component of the motion towards our line of sight (i.e. $\le90^o$)
our simulation reproduces remarkably well the Chandra observations of the
arcs.

\section{THE ``AXIAL" JET !}

 We now turn to the jet like feature whose 
symmetric bisection of the arcs discussed above
has prompted immediate
identification with its rotation axis. The fact that the projection on the
sky-plane of the rotational axis as derived from the polarization data 
(with a 90$^o$ shift) is
also in agreement has been interpreted as further and strong evidence of
the above supposition. In fact, Lai et al. (2001) go so far as to say
``............ If the jet originates from the pulsar magnetosphere, as
seems likely, it is most natural to associate the jet axis with the pulsar
spin axis." The alignment of the jet with the observed proper motion for
Vela seems also to have prompted an association with, and raised hopes of
an explanation for the space velocity of the neutron star. The fact that a
similar jet was already seen in the Crab, also aligned with the derived
P.A. of the spin axis, and its proper motion, heralds an incipient
industry as more X-ray images of radio pulsar nebulae become available.

A radio pulsar is not an accreting object like an x-ray pulsar, a black
hole binary, or an AGN. In our view, it is not only unlikely, but
unphysical to expect an actual jet along the rotation axis of a radio
pulsar. There is an immense amount we don't know about the magnetospheres
of radio pulsars, but one thing we have known for over three decades is
that the particle emission is along the magnetic axis (Goldreich \& Julian 1969),
that is often at
large angles to the rotation axis. The only connection that there can
possibly be between the X-ray jet and the rotation axis is one of
projection on the sky-plane as we shall argue below. 

\begin{figure}
\epsfig{file=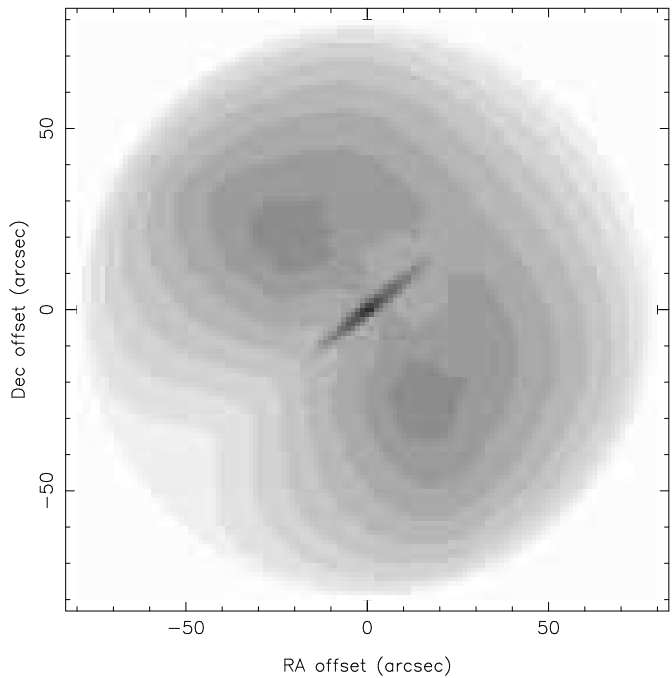,width=3.5in,angle=-90}
\caption[]{
Gray-scale plot showing the distribution of the expected
X-radiation corresponding to the `jet' and diffuse 
components around the pulsar.
The RA \& Dec. offsets are with respect to the pulsar position.
The particle flow causing these components of the
 X-radiation is assumed to be in the form of
an elongated fan beam (with angular spreads of $\pm5^o$
and $\pm70^o$ in the longitudinal \& latitudinal directions respectively)
centred around the rotating magnetic axis vectors associated 
with both the poles.
The particles are allowed to suffer angular deviations
only in the latitudinal
direction and are assumed to be confined to the $\pm70^o$ spread of the
fan beam for consistency.
 All the values of relevant angles associated with our viewing geometry
are as estimated from the best fit solution  described in the 
earlier section.
}
\end{figure}

In our picture of the
relativistic particle beams leaving the magnetic poles and proceeding
ballistically outwards to the cavity walls, we can ask what if any
deviations they are likely to suffer. The only field along their
trajectory through the cavity is the toroidal field (that they are
carrying out) and that later gets compressed and strengthened at the wall.
This toroidal field will, over most or all of the trajectory, be
perpendicular to the path of the particles, and, of course, the rotation
axis. Depending on the spread of their energies, it should not be
surprising if a small fraction of the particles acquired a spread of
velocities in the latitudinal direction making their synchrotron radiation
visible to an observer when their motion is tangential to the sight line.

One should, as a consequence, expect to receive radiation from these
particles when the projection of the magnetic axis coincides with that of
the rotation axis, precisely as in the case of the radio pulse, but now
over a larger range of angles in latitude. Wherever the observer, the
apparent jet will appear along the minor axis of the projected ellipses of
the arcs but the extent over which it is visible will depend on the spread
of particle velocities, and the angles the beams from the two poles make
to the line of sight. These particles must also reach the cavity walls and
will then create a diffuse glow around the arc regions but with a greater
spread, exactly as seen in the Chandra image. We have assessed the spread
of this weak fan beam from the size of the diffuse glow, and from the
poor or non-visibility 
of the corresponding radiation from the beam of the other
magnetic pole. We estimate a spread of about $\pm$70$^o$ from the observed
extent of the jet and have
used this value in
our simulation (fig. 2) of the jet and the associated diffuse components.

 If this interpretation of the ``jet" is correct, it has a
major implication for the picture of particle flow from the pulsar to the
nebula. In section 4 when attempting to explain the formation of the X-ray
arcs, we made the assumption that the particle beam separated from the
polar field lines well within the light cylinder and before the effect of
any sweep-back. 
The justification for this assumption is now seen as the absence of
any misalignment between the apparent jet (defined by the trajectory of
particles radiating towards us) and the projected rotation axis,
the very observation that prompted the physical misinterpretation of the
jet referred to earlier.

The total picture that emerges is as follows. 
There is a cavity of radius of order $10^{17}$ cms,
which we can presume was created by the dipole radiation originally of
much higher frequency than the present 11 Hz. Inside is a double cone of
half angle $\alpha\sim$70$^o$,
along which there is a relativistic particle flow in
straight lines with a drift time of $\sim$0.1 year before they encounter the
``shock". In addition, there is a low density of particles separating from
the cones but whose trajectories are in planes containing the rotation
axis. Any observer will see radiation only from those particles in the
plane containing the observer and the rotation axis and we predict that
this radiation should be highly linearly polarized with the electric
vector parallel to the rotation axis.

\section{ DISCUSSION}

In the long march towards the elucidation of
the mysterious ways of pulsars, now numbering around a thousand, a few
special ones have taught us more than most of the rest put together.
The Vela Pulsar is one such, and played an important role in several ways
within months of its discovery in 1968. The superb capabilities of the
Chandra telescope have put this
pulsar in the limelight again by providing a spectacular image in X-rays
of the surrounding nebula. Its form and proportions, reminiscent of
pre-Columbian pectoral ornaments are loaded richly with information about
many aspects. To begin with, the symmetry of the nebula has provided
compelling evidence for the identification of the P.A. of the rotational
axis of the pulsar. The near precise orthogonality of this P.A. to that
inferred with certain assumptions from polarization measurements, has had
important implications. We support the finding of Lai et al. (2001) and
Helfand et al. (2001) that the dominant polarization mode in the Vela
Pulsar has the electric field orthogonal to the magnetospheric field. We
find this is in accordance with the mechanism of Luo \& Melrose (1995)
that appears to explain three important characteristics of the radio
radiation. They are the high brightness temperatures, the dominant
polarization mode, and the observed sweep of the P.A. across the pulse,
the last clearly identifying the input signal to the maser as curvature
radiation from the field lines in the polar neighbourhood. Since the first
and third characteristics also apply to the parallel polarization mode to
which the radiation occasionally flips, as seen in many pulsars, we
propose that the same amplification mechanism must also be operative at
times in the other mode. If the Luo \& Melrose mechanism can in fact
operate in both modes, the explanation of any type of elliptic
polarization becomes a matter of detail, a topic which we shall address
later.

Noting the clear separation of the X-ray nebular
emission into two elliptic arcs symmetrically located with respect to the
inferred rotation axis, we propose that they are the traces of the two
particle-beams from the magnetic poles
on the walls of the cavity $--$ the shocked region. We explain
the visible extent of the arcs as arising simply from the spread of the
pitch angles at the shock front. We also explain the alignment with the
rotational axis of the jet like feature bisecting the arcs as simply a
projection effect on the sky plane. We attribute its visibility to a
latitudinal spread of the velocities of a small fraction of the particles
in the beam, an explanation strongly supported by the presence of diffuse
emission around the arcs. We present a simulation of the expected
X-radiation around the pulsar based on the above model for the arcs,
the jet, and the diffuse glow around both (see fig. 3),
and find gratifying agreement
with the Chandra observations\footnote{
Image available at\\ http://chandra.harvard.edu/photo/cycle1/vela}.
 We predict that the polarization of the jet
feature will be linear and parallel to it, and claim that there can be no
physical jet along the rotational axis of the pulsar.

\begin{figure}
\epsfig{file=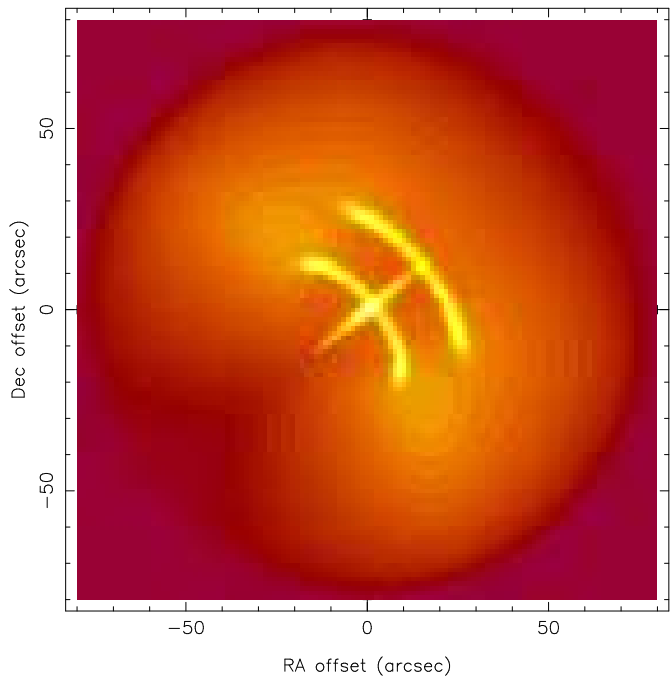,width=3.5in,angle=-90}
\caption[]{A result of our simulations as in Fig. 2, 
but now including also the arc components.
The colour code used here is roughly similar to that in the Chandra image
referred to in the text.}
\end{figure}

Note that the `jet' is really an apparent one.
There are at least 3$\times$10$^7$ input bunches of particles along any line from
the pulsar to the cavity wall, and even the slightest dispersion in velocity
would smooth this out to a uniform flow at a very short distance from the pulsar.  
Even in the absence of velocity dispersion, the finite gamma of the bunch
combined with the distance over which it is radiating will result in smearing.
The radiation will thus 
appear continuous and be a part of the unpulsed fraction in X-rays,
the fraction depending on the 
extent of the jet/nebula from which radiation is collected.

In our modelling of the pair of arcs as traces of the two polar beams
on a cavity wall (e.g. as illustrated in Fig 1), we find that an assumption
of equatorial symmetry does not fit the observations well.
As already mentioned in an earlier section, the best fit
$r_1$ is significantly different from $r_2$ (i.e. $(r_2/r_1)\approx$1.35). 
Note that, in our model, $r_1$ \& $r_2$ represent the implied distances to the
``wall'' from the star along the `rotating vectors' associated with the
near and the opposite poles respectively. 
The arcs provide us important information including about  
the otherwise `unseen' pole.
Significantly improved fits are obtained when unequal values of
$\alpha_1$ \& $\alpha_2$ 
(the half angles of the polar cones associated with the two poles) 
are allowed in the model. 
The corresponding $r_1$,$r_2$ are unequal again. 
Even better fits are obtained if $\beta$ is not constrained by the radio
observations. But interestingly, 
the implied value of $\zeta$ (=$\alpha_1 + \beta$), the angle between
the rotation axis and our line of sight, is about the 
same as in the other cases.
Further detailed modelling, than 
has been possible presently, may provide better
estimates of the above parameters and
clues about the size as well as the shape of the `cavity'.

Inequality between $\alpha_1$ \& $\alpha_2$ has
implications, particularly for certain acceleration 
mechanisms for the origin of pulsar velocities. 
For example,
the `rocket' mechanism of Harrison \& Tademaru (1975) does necessarily
require such an inequality amounting to a tilted and offset dipole.

Other
possible/plausible implications are the following.

i) If the dominant radio polarization mode is indeed the `orthogonal' mode, then
any analysis based on the assumption that the observed central polarization 
P.A. is the same as the P.A. of
the rotation axis needs to be reviewed. This applies, for example,
to the comparisons
of the proper motion directions of pulsars with the orientations of their
rotation axis.
In the work of Deshpande et al. (1999), 
they allowed for
mode ambiguity in cases where emission of both modes is observed, and
have assessed the distribution of the proper motion direction (relative to
the rotation axis P.A.) as shown in their Fig 1b. 
We argue that the required revision will amount to simply replacing the
relative angle by its complement. Since no particular preference was found 
for any relative angle value, the conclusions of Deshpande et al. remain
unaltered. 

ii) Given the similarities observed
between the  morphologies of the surrounding nebulae 
as well as other properties of
the Vela \& Crab pulsars, it would not surprise us if
a similar arc structure is revealed around the Crab pulsar by
observations with improved spatial resolution.

\begin{acknowledgement}
 We would like to thank Don Melrose, Qinghuan Luo and Mark Walker 
 for many useful discussions, and Richard Dodson for kindly providing
 information prior to publication.
\end{acknowledgement}

\end{document}